\documentclass[fleqn,usenatbib]{mnras}

\usepackage{newtxtext,newtxmath}
\usepackage[T1]{fontenc}

\DeclareRobustCommand{\VAN}[3]{#2}
\let\VANthebibliography\thebibliography
\def\thebibliography{\DeclareRobustCommand{\VAN}[3]{##3}\VANthebibliography}

\usepackage{graphicx}	% Including figure files
\usepackage{amsmath}	% Advanced maths commands
\usepackage{booktabs}
\usepackage{tikz}

\title[Exoplanet Ring-Moon Cycles]{On The Applicability of Ring-Moon Cycles to Exoplanets}

\author[I.E. Ward]{Isabella E. Ward,$^{1}$
and Matija \'Cuk$^{2}$\thanks{E-mail: mcuk@seti.org (M\'C)}
\\
$^{1}$Tufts University, Medford, MA, USA\\
$^{2}$SETI Institute, Mountain View, CA, USA
}

\date{Accepted XXX. Received YYY; in original form ZZZ}

\pubyear{\the\year{}}

\begin{document}
\label{firstpage}
\pagerange{\pageref{firstpage}--\pageref{lastpage}}
\maketitle

\begin{abstract}

The presence of rings and moons around exoplanets is likely to be one of the next great discoveries in exoplanet research. Using theories developed for the Solar System, we explore the possibility of coupled ring-moon cycles around exoplanets and what these processes mean for the observability of these features. Around Neptune- and Earth-like planets, we find that ring-moon cycles are capable of producing long-lived rings of comparable and greater relative transit depths than Saturn's. In multi-planet systems, secular spin-orbit resonances can provide the necessary planetary obliquity for these rings to contribute noticeably to transit lightcurves. We model the geometry of a ring's cross-section at various angles in comparison to the cross-section of a transiting planet to determine whether the ring may be detectable during the planet's transit. Ringed planets have also been proposed as an alternative to inflated planetary radii seen in transit, leading to abnormally low observed densities. Ring-moon cycles can produce late-forming and sometimes long-lived rings that can have the potential of explaining at least some of these observations. We also discuss some inconsistencies in the calculation of exoplanet oblateness due to rotation that we have come across in the course of this work.

\end{abstract}


%%%%%%%%%%%%%%%%%%%%%%%%%%%%%%%%%%%%%%%%%%%%%%%%%%

%%%%%%%%%%%%%%%%% BODY OF PAPER %%%%%%%%%%%%%%%%%%

\section{Introduction}

Rings and moons are common across the Solar System, with hundreds of moons known to be orbiting planets, Trans-Neptunian Objects \citep{Noll08}, and asteroids \citep{Margot15}, while ring systems are found around the four gas giant planets, the dwarf planet Haumea, and multiple Centaurs \citep{sicardy2018}. Even planets that do not currently have rings have been proposed to have had them in the past. Massive rings created from impacts on Earth and Mars are theorized to have led to the formation of the Moon \citep{Canup23} and Martian satellites Phobos and Deimos \citep{Rosenblatt16}. Rings and moons not only are both features common among Solar System objects, but they share a close evolutionary link. Moons are thought to often form from rings \citep{Charnoz11, crida2012}, and this process may be currently ongoing in Saturn's rings \citep{Charnoz10}. On the other hand, rings can form from a tidal \citep{Wisdom22} or collisional \citep{Teodoro23} break-up of satellites. In some cases, both of these processes may occur within the same system; around Uranus, for example, part of the ring system present is theorized to have originated from a disrupted moon \citep{showalter2006} that may re-accrete in the future \citep{french2012}. In this paper we will address cases in which both of these processes may have happened multiple times around the same planet.

\citet{HM17} describe a cyclical process through which material from planetary rings accrete into satellites, which may eventually migrate inward toward the planet to be broken up into new rings, restarting the process. 
Such a cycle was initially presented as an explanation of the origin of the Martian moon Phobos, which is expected to be torn into a ring in less than 100 Myr as it migrates closer to Mars \citep{Black15}. Observing Phobos at the very end of its life by chance appears unlikely, but \citet{HM17} explain this apparent anomaly by Phobos being only the latest in a sequence of inner Martian moons that have been destroyed into rings, only so that the next, less massive moon can form at approximately the same location and continue this cycle. 

\citet{HM19} expand the ring-moon cycle theory beyond the case of Mars, and find that planets can be classified into three groups on the basis of potential for ring-moon cycles (see also Sec. \ref{non-synch}). Planets with Roche limits well within the synchronous orbit, such as Mars, are firmly in the regime where these cycles are possible. Planets with Roche limits beyond synchronous orbit, such as Saturn, can only have outward-migrating moons. \citep{HM19} find Uranus to be in a regime where the plausibility of ring-moon cycles depends on the specific properties of the rings and moons. Here we are assuming that there is a single Roche limit for all ring material, but in reality different parts of the ring  system can have different compositions (and therefore densities), and also ring particles have widely different porosities, which all affect the concept of the Roche limit \citep{Tiscareno13}. 

Uranus's ability to maintain ring-moon cycles depends on the magnitude of ring torques acting on its satellites. If the tidal torques from interactions with the planet on a Uranian satellite exceed the Lindblad torques from interactions with the ring, then the satellite will be pulled inward until it is pulled apart into a new ring \citep[][]{HM19}. Although the moon Miranda has been proposed by \citet{HM19} as a possible result of such cycles, more recent studies find that ring-moon cycles are more likely to have produced only the smaller inner moons closer to Uranus \citep{cuk2022}. Specifically, past existence of rings as a part of ring-moon cycles can explain the orbital resonance between moons Belinda and Perdita. Ring-moon interactions would make Belinda migrate outward and form the resonance, while tides have the opposite effect and cannot make the two moons migrate convergently.

Likely existence of ring-moon cycles around Solar System planets suggests that such processes may occur around exoplanets as well. In order to confirm this, we must find these features first: thus far, no exoplanetary rings or satellites have been directly detected \citep{aizawa18, ohno22}, but processes like planetary transits show promise that we may be able to detect rings dependent on the orientation of the planet \citep{barnes2004, Heising15, Heller18}. Edge-on ring systems are less likely to be detected, but planets with significant obliquity may have rings that are observable during transits. While despun planets are expected to have zero obliquity, in multi-planet systems secular resonances can push planets towards greater obliquities. We explore the plausibility of known exoplanets experiencing ring-moon cycles, and whether the rings created in these cycles could be observable based on the overall parameters of the system. Successful observations would provide insight into exoring populations, aid in the study of exoplanet interiors \citep{schlichting2011}, and, given the link between rings and moon formation, could lead closer to the discovery of exomoons as well. If exorings were to be directly detected, this may offer us invaluable information of the operation of ring-moon cycles beyond the Solar System.

We first investigate the plausibility of ring-moon cycles around known, likely non-despun exoplanets and whether they could have optically thick ring systems. Next, we explore the same in multi-planet despun systems, with the addition of searching for potential secular spin-orbit resonances that may tilt the planet's spin axis, which may allow rings to be observable in transit lightcurves. Finally, we explore the observability of such rings by analyzing their contributions to planetary cross-sections for different spin-axis orientations.

\section{Non-Synchronous Planets}\label{non-synch}

The ability of a planet to maintain ring-moon cycles is heavily dependent on Roche limits, which determine the threshold at which satellites will begin to be tidally disrupted by the planet they orbit, and synchronous orbits, where the rotational period of the planet is equal to the orbital period of the satellite. The location of the synchronous orbit is vital to determining which planets may go through ring-moon cycles. Planets with rapid rotation rates have closer synchronous orbits, and as tidal dissipation drives satellites away from the synchronous distance \citep{murray1999}, many close-in moons tend to be driven away, as opposed to being pulled back inward to be broken up. In this section we identify non-despun Neptune-like planets, which we assumed to have Neptune-like relatively slow rotation \citep[with some observational support;][]{lammers2024, Price25}, explore their ability to maintain ring-moon cycles, and discuss whether the rings made from such cycles could potentially be optically thick.

\subsection{Planet Selection}
We investigate a sample of Neptune-like (3-5 R$_{Earth}$) transiting exoplanets, starting with all known planets within this radius range with orbital periods greater than 25 days. We thus begin with a total sample of 164 planets. Such planets, if not despun by their host star, may spin at such a rate that allows for ring-moon cycles to occur. To identify which planets are likely not to be despun, we calculate the change in the rotation rate of a planet as \citep{murray1999}
\begin{equation}
\label{eq:1}
    \dot \Omega_p=-sign(\Omega_p - n)\frac{3k_{2p}}{2\bar C_pQ_p} \left( \frac{M_{star}}{M_p}\right)\left(\frac{R_{star}}{a}\right)^3n^2
\end{equation}

\noindent where $\Omega_p$ is the planet's rate of rotation, $k_2$ is the planet's Love number, $\bar C_p \equiv C_p / {M_pR_p^2}$ is the dimensionless moment of inertia of the planet, and $Q_p$ is its tidal quality factor; $m_s$ and $m_p$ are the masses of the star and planet respectively, while $R_s$ and $a$ are the planet's radius and semimajor axis. Finally, $n$ denotes the planet's mean motion. For a Neptune-like planet, we assume $k_2 = 0.1$ and $Q = 10,000$. These values are based on continuing survival of Triton on a retrograde orbit around Neptune \citep[e.g.][]{chyba1989}; here we assume classic equilibrium tides and do not consider resonant tides which can be present in gas giants \citep{Fuller16}. We also assume a spin period of 60,000 seconds (or approximately 16.67 hours), allowing us to estimate $\Omega_p = 10^{-4}$~rad~s$^{-1}$. In this case, since the spin period is assumed to be much less than the orbital period for all planets, the sign is always negative. This is inversely proportional to the planet's spindown time, such that $T = \frac{\Omega_p}{\dot \Omega_p}$.

\begin{figure}
	\includegraphics[width=\columnwidth]{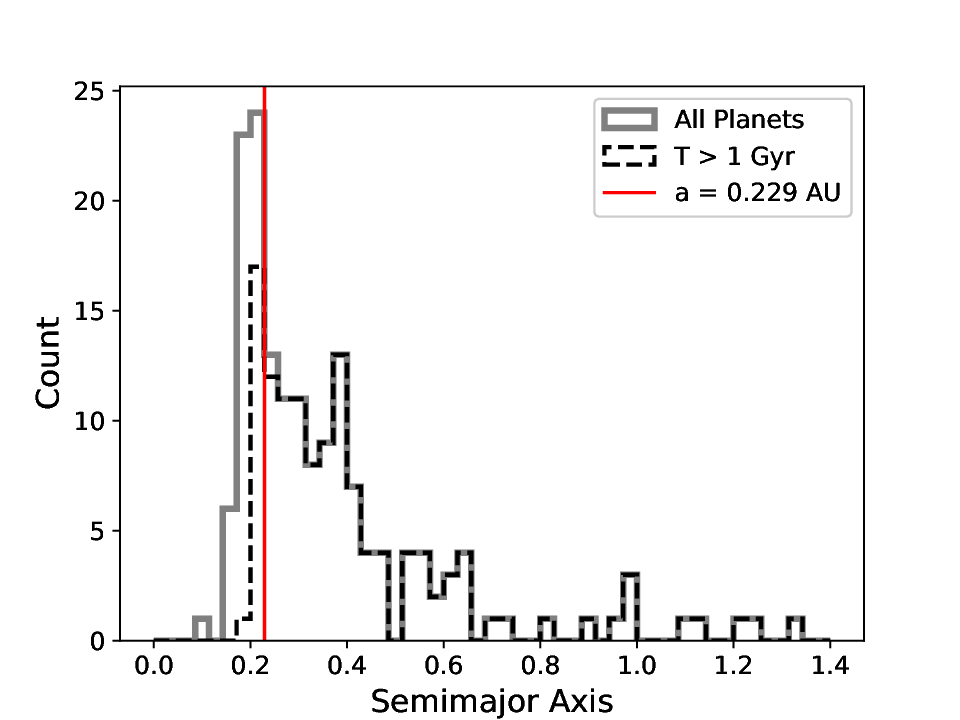}
    \caption{The total sample of planets sorted by semimajor axis compared to the number of planets with despinning timescales of over 1 Gyr. All planets in the sample with semimajor axes over 0.229~au have despinning timescales beyond this threshold.}
    \label{fig:non_despun_hist}
\end{figure}

From the initial sample of planets, we then select those for which $T > 1$ Gyr. These planets have a sufficiently long spindown time to not yet have been despun by their host stars. A comparison of such planets as a portion of the entire sample can be found in Figure~\ref{fig:non_despun_hist}.

We also confirm that all the selected planets are within their respective ice lines using
\begin{equation}
    y = \left( \frac{R_{star}}{R_{sun}} \right)^2 \left( \frac{T_{eff,star}}{T_{eff,sun}} \right)^4 \left( \frac{3\ {\rm au}}{a} \right)^2
\end{equation}
where $y$ denotes the planet's relative position to the ice line, which sits at y = 1. 
All planets yield y > 1, setting them within the ice line for their system. This information is used to estimate the density of the ring material (i.e. rock vs. ice). 

\subsection{Roche Limits, Synchronous Orbits, and Lindblad Torques}

Before investigating hypothetical ring-moon cycles, we must first verify that the planets in question have the potential to host such cycles at all. \citet{HM19} identify three evolutionary regimes for moons accreted from ring material, two of which are compatible with ring-moon cycles. To explore these, we first must understand the Roche limit, synchronous orbits, and Lindblad torques. These can be complex concepts with varying definitions; here we will mostly follow the approach of \citet{HM19} as we are primarily interested in ring-moon cycles. 

The Roche limit describes the distance at which a satellite must remain away from a planet in order to remain intact, assuming that it is held together by gravity rather than cohesive forces  \citep{murray1999, Holsapple08}. Upon passing within the Roche limit, the self-gravity of the satellite no longer dominates over the gravity of the planet, and material on the satellite's surface begins to get pulled away.

The Rigid Roche Limit assumes that the satellite is a rigid sphere, in which case the semimajor axis of the Roche limit can be calculated as \citep[e.g][]{HM19}
\begin{equation}
    a_{RRL} \approx 1.442R_p \left( \frac{\rho_p}{\rho_s} \right)^{\frac{1}{3}}
\end{equation}
where $R_p$ is the radius of the planet, $\rho_p$ is the planet's bulk density, and $\rho_s$ is the satellite's bulk density.

However, satellites are often more similar to a prolate spheroid than a sphere, and in this case effect of self-gravity is lessened near the equator as material at the surface resides farther from the center of mass than material in other regions. The shape of the satellite can be roughly approximated to be close to hydrostatic equilibrium, which implies that (over long timescales) the moon behaves like a fluid \citep[or at least as a ``rubble pile'';][]{Sharma09, Madeira23}. Here, the Fluid Roche Limit (FRL), which is set farther from the planet than the RRL, is appropriate. This is calculated as \citep[e.g][]{HM19}
\begin{equation}
    a_{FRL} \approx 2.456R_p \left( \frac{\rho_p}{\rho_s} \right)^{\frac{1}{3}}
\end{equation}

Another important factor is the location of the synchronous orbit, where the orbital period of a satellite matches the rotational period of its planet. The semimajor axis of the synchronous orbit is given by \citep[e.g][]{HM19}
\begin{equation}
    a_{synch} = \left( \frac{G(M_p + M_s)T_p^2}{4\pi^2} \right)^\frac{1}{3}
\end{equation}
where $M_p$ and $M_s$ are the masses of the planet and satellite, respectively, and $T_p$ is the planet's rotational period. The location of the satellite's orbit relative to the synchronous orbit affects whether tidal torques from the planet pull the satellite inward toward the planet or push it outward, with the former occurring within the synchronous orbit and the latter occurring outside it.
Finally, when a satellite forms, it gravitationally interacts with the material of the ring it formed from. The material left in the ring is gravitationally attracted to the satellite, and the resulting density perturbations cause torques between the satellite and ring called Lindblad torques that migrate the satellite away from the planet. The farthest orbit a satellite can migrate to due to these torques is one for which the most distant first-order Lindblad resonance of the satellite (2:1) is at the outer edge of the ring, and has a semimajor axis of \citep{HM19}
\begin{equation}
    a_{Lind} = 4^\frac{1}{3}a_{FRL}
\end{equation}

Together, all of these values can be used to classify a given planet's system into one of three evolutionary regimes: the "Boomerang" regime, where satellites initially migrate outward but later begin to move back toward the planet; the "Slingshot" regime, where they migrate outward and do not change direction; or the "Torque-Dependent" regime, where either of the previous options can be the case depending on the torques acting on the satellite \citep{HM19}.

In the "Boomerang" regime, the maximum semimajor axis a moon can reach due to Lindblad torques is smaller than that of the synchronous orbit with the planet; in the "Torque-Dependent" regime, the opposite is true, but the semimajor axis of the synchronous orbit is beyond the Fluid Roche Limit. Planets within either of these regimes may successfully maintain ring-moon cycles, although this is not always the case for the latter. The "Slingshot" regime, where $a_{synch}$ is closer to the planet than $a_{FRL}$, is the only one in which the satellite is always driven away from the planet. 

We calculate $a_{RRL}$, $a_{FRL}$, and $a_{Lind}$ using an assumed density of 2400 kg m$^{-3}$ \citep[based on densities of small S-type asteroids;][]{Cheng23} for ring material since all of our planets are within the ice line. We again use a planetary spin period of 60,000 seconds for $a_{synch}$. All of our selected planets thus fall into the "Boomerang" regime, implying potential for ring-moon cycles to occur throughout our entire selection.

\subsection{Ring-Moon Lifetimes and Equivalent Ring Thickness}
\label{sec:ring_life}

The likelihood of observing a ring created during ring-moon cycles depends heavily on what fraction of time the rings are present and their thickness. We can estimate both of these aspects by investigating the lifetimes of hypothetical ring-moon cycles around the sample of planets we have acquired. In this section, we analyze the lifetimes of rings and satellites formed during ring-moon cycles and investigate their observability by calculating a quantity that is arguably related to the ring optical depth late in the planet's lifetime.

The lifetime of a planetary ring that is viscously spreading and accreting into moons at the Roche Limit is estimated by \citet{crida2012} as:
\begin{equation}
    \tau_{ring} \approx 0.0425 \left( \frac{M_{disk}}{M_p} \right)^{-2} \sqrt{\frac{4\pi^2}{GM_p}a_{FRL}^3}
\end{equation}
and the lifetime of a satellite against tidal decay is given by \citep{HM19}
\begin{equation}
    \Delta t = \frac{2Q_p}{39M_sR_p^5k_2} \sqrt{M_p/G}
\end{equation}

\begin{figure}
    \includegraphics[width=\columnwidth]{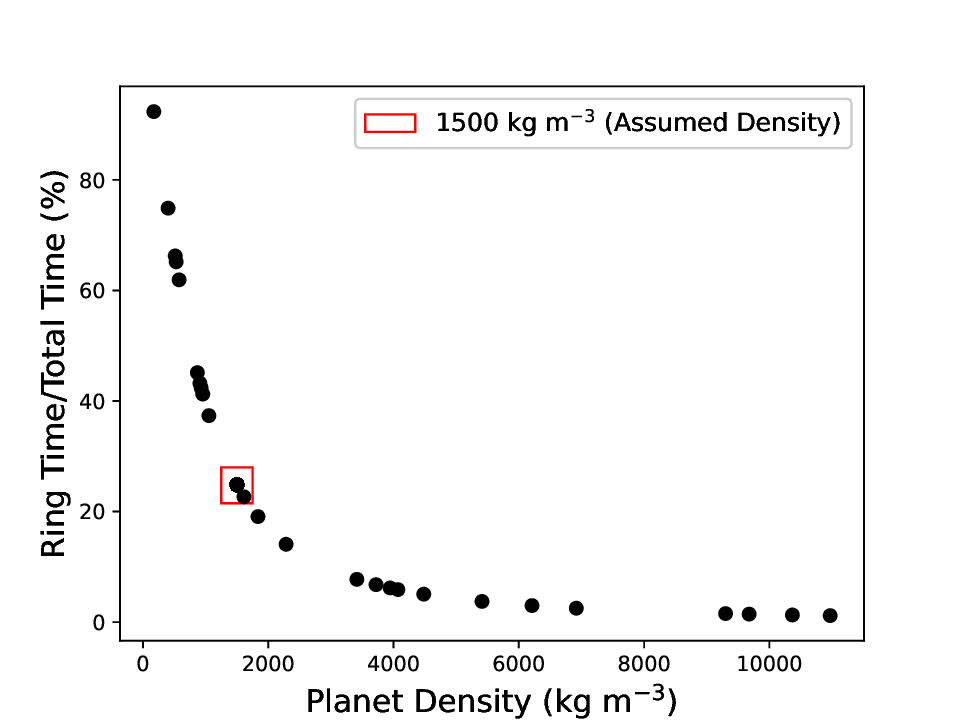}
    \caption{The amount of time rings would be present for each non-despun planet (represented as black points) out of overall time as a function of planet density. The red box marks the 1500 kg m$^{-3}$ density assumed for all planets in the sample without defined masses, which subsequently all result in the same ''ring time'' percentage.}
    \label{fig:density_ringtime}
\end{figure}

For each planet, we calculate the lengths of alternating ring and moon lifetimes until the ring lifetime meets or exceeds 1 Gyr. We start with a ring mass $M_{ring} = 10^{-4}M_p$ \citep[equal to the typical mass of the whole satellite system;][]{canup2006} and, following estimates from \citet{HM17}, we assume that each next generation moon has 20\% of the mass of the preceding ring. The following ring is then set to be the same mass as the moon that created it. We keep record of the length of time for which the planet has rings and for which it has moons, then compare the ``ring time'' within the overall timescale. The overall timescale encompasses the total time period through all cycles plus the lifetime of the satellite that would accrete after the first ring lasting 1 Gyr or more. Figure~\ref{fig:density_ringtime} shows this ratio as a function of planet density. Note that a large fraction of planets without known masses are in the red box with assumed density of 1500~kg~m$^{-3}$, based on the idea they are Neptune-like bodies. Such planets would have rings more than 20\% of the time. Many of the points with high densities may not correspond to real planetary parameters, as mass measurements can be very uncertain.

\begin{figure*}
    \includegraphics[width = \textwidth]{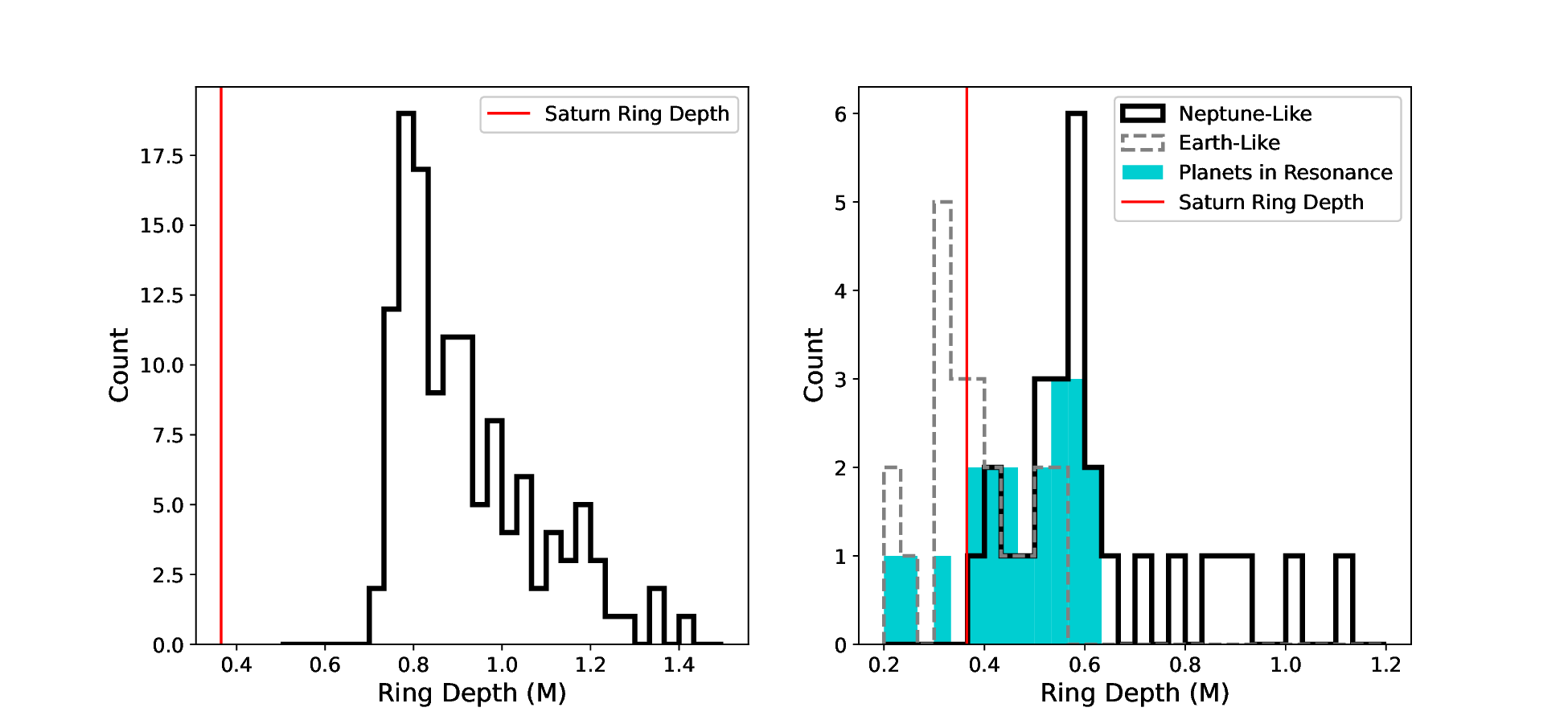}
    \caption{The number of planets meeting or exceeding an equivalent ring thickness threshold of 0.3651 meters, or the approximate Saturn ring thickness $d$ as calculated. On the left, we consider all non-despun planets; on the right, despun planets in multi-planet systems, with specific attention to planets in potential spin-orbit resonances.}
    \label{fig:ring_depths}
\end{figure*}

The optical depth of a ring represents how much light is blocked out by ring material, and is a function of the transparency of the ring. This in turn is dependent on the ratio between the cross-sectional area of ring particles and the surface area of the entire ring space. To identify a factor by which we can determine the observability of the ring without knowing the radius of ring particles, we calculate a factor of the transparency: the ''equivalent thickness'' of the ring, $d$, given by
\begin{equation}
    d = \frac{M_{ring}}{\pi\rho_{ring}(a_{FRL}^2 - R_{planet}^2)}
\end{equation}
where $\rho_{ring}$ is the same 2400~kg~m$^{-3}$ as previously used. To compare, we perform the same calculation to estimate the equivalent thickness of Saturn's rings, treating all the different rings as one system. Based on a mass of $1.54 \times 10^{19}$ kg \citep{iess2019}, this yields an equivalent thickness of 0.365 m. Given that Saturn is known to have optically thick rings, we set this value to be the threshold for optically thick rings within our sample.  The exact optical depth depends, of course, on the size distribution of particles. Here, we conclude that the total ``uniform slab'' thickness of the rings compared to the main rings of Saturn may still be optically thick $\tau >> 1$ and therefore potentially observable during transits. The left plot of Figure~\ref{fig:ring_depths} shows the distribution of equivalent ring thickness across non-despun planets.

\section{Synchronous Planets in Spin-Orbit Resonances}

\begin{table*}
\caption{We show the period, mass, and radius of each planet in the ten sample multi-planet systems. Parameters were collected from the NASA Exoplanet Archive \citep[https://exoplanetarchive.ipac.caltech.edu/, accessed 9 June 2025, see also][]{Christiansen25}. }
\begin{tabular}{lrrr}
\toprule
\toprule
Name & Period (Days) & Mass & Radius \\
\midrule
Kepler-11 \\
\midrule
b & 10.303900 & 1.90 & 1.80 \\
c & 13.024100 & 2.90 & 2.87 \\
d & 22.684500 & 7.30 & 3.12 \\
e & 31.999600 & 8.00 & 4.19 \\
f & 46.688800 & 2.00 & 2.49 \\
g & 118.380700 & 25.00 & 3.33 \\
\midrule
Kepler-33 \\
\midrule
b & 5.667930 & 5.27 & 1.74 \\
c & 13.175620 & 8.94 & 3.20 \\
d & 21.775960 & 41.80 & 5.35 \\
e & 31.784400 & 17.73 & 4.02 \\
f & 41.029020 & 24.22 & 4.46 \\
\midrule
Kepler-80 \\
\midrule
b & 7.052460 & 6.93 & 2.67 \\
c & 9.523550 & 6.74 & 2.74 \\
d & 3.072220 & 6.75 & 1.53 \\
e & 4.644890 & 4.13 & 1.60 \\
f & 0.986787 & 1.77 & 1.21 \\
g & 14.645580 & 1.44 & 1.13 \\
\midrule
Kepler-186 \\
\midrule
b & 3.886791 & 1.23 & 1.07 \\
c & 7.267302 & 1.95 & 1.25 \\
d & 13.342996 & 2.75 & 1.40 \\
e & 22.407704 & 2.05 & 1.27 \\
f & 129.944100 & 1.60 & 1.17 \\
\midrule
Kepler-223 \\
\midrule
b & 7.384490 & 7.40 & 2.99 \\
c & 9.845640 & 5.10 & 3.44 \\
d & 14.788690 & 8.00 & 5.24 \\
e & 19.725670 & 4.80 & 4.60 \\
\bottomrule
\end{tabular}
\begin{tabular}{lrrr}
\toprule
\toprule
Name & Period (Days) & Mass & Radius \\
\midrule
Kepler-305 \\
\midrule
b & 5.487000 & 10.50 & 3.60 \\
c & 8.291000 & 6.00 & 3.30 \\
d & 16.740000 & 6.20 & 2.76 \\
e & 3.205380 & 5.78 & 1.79 \\
\midrule
Kepler-758 \\
\midrule
b & 12.109710 & 4.16 & 2.48 \\
c & 4.757940 & 4.83 & 1.69 \\
d & 20.496620 & 2.60 & 2.12 \\
e & 8.193472 & 3.58 & 1.53 \\
\midrule
TOI-1246 \\
\midrule
b & 4.307440 & 8.10 & 3.01 \\
c & 5.904137 & 9.10 & 2.45 \\
d & 18.654874 & 5.40 & 3.43 \\
e & 37.925480 & 14.50 & 3.51 \\
\midrule
TOI-2076 \\
\midrule
b & 10.355235 & 3.73 & 2.39 \\
c & 21.015327 & 55.00 & 3.69 \\
d & 35.125686 & 29.00 & 3.43 \\
e & 3.022345 & 2.49 & 1.36 \\
\midrule
TRAPPIST-1 \\
\midrule
b & 1.510826 & 1.37 & 1.12 \\
c & 2.421937 & 1.31 & 1.10 \\
d & 4.049219 & 0.39 & 0.79 \\
e & 6.101013 & 0.69 & 0.92 \\
f & 9.207540 & 1.04 & 1.05 \\
g & 12.352446 & 1.32 & 1.13 \\
h & 18.772866 & 0.33 & 0.76 \\
\bottomrule
\end{tabular}
\label{tab:systems_table}
\end{table*}

If they form similarly to Earth \citep{Kokubo10, Lock18, Canup23}, Super-Earths likely have rapid primordial spins that would typically result in moons migrating away from the planet. Despinning from host stars may allow some  Super-Earths to maintain ring-moon cycles. For us, a particularly interesting subset of such cases is planets in multi-planet systems, where secular spin-orbit resonances among the planets \citep[cf. ][]{Ward04, Hamilton04} can induce obliquities even in tidally despun planets \citep{Millholland19, Millholland2024}. In this case, it could be possible to observe the rings created in such cycles if the planet transits its host star, as the cross section of the rings will block additional light during transit. In this section we identify potential secular spin-orbit resonances among planets in multi-planet systems and evaluate whether these planets could develop visible rings from ring-moon cycles.

\subsection{Identifying Secular Resonances}

\begin{table*}
\begin{tabular}{lrrrr}
\toprule
\toprule
Planet & Nodal Precession & Spin Precession & Timescale (Gyr) \\
\midrule
Kepler-11 \\
\midrule
b & -0.006019 & -0.010482 & 103.320737 \\
c & -0.006019 & -0.011549 & 24.172007 \\
e & -0.000561 & -0.000779 & 612.463241 \\
\midrule
Kepler-33 \\
\midrule
b & -0.050831 & -0.062975 & 457.546503 \\
c & -0.006156 & -0.011155 & 153.705816 \\
\midrule
Kepler-80 \\
\midrule
b & -0.066952 & -0.072739 & 24.673599 \\
c & -0.025308 & -0.029539 & 28.476441 \\
e & -0.082461 & -0.114424 & 19.180957 \\
e & -0.066952 & -0.114424 & 14.018880 \\
\midrule
Kepler-186 \\
\midrule
c & -0.019304 & -0.029876 & 10.612079 \\
\midrule
Kepler-223 \\
\midrule
b & -0.045066 & -0.063362 & 9.541115 \\
c & -0.023697 & -0.026734 & 22.780108 \\
\bottomrule
\end{tabular}
\begin{tabular}{lrrr}
\toprule
\toprule
Planet & Nodal Precession & Spin Precession & Timescale (Gyr) \\
\midrule
Kepler-305 \\
\midrule
c & -0.037096 & -0.044768 & 6.950763 \\
d & -0.005392 & -0.005439 & 5879.918566 \\
\midrule
Kepler-758 \\
\midrule
d & -0.002733 & -0.002963 & 5199.263244 \\
e & -0.015025 & -0.020847 & 179.003305 \\
\midrule
TOI-1246 \\
\midrule
c & -0.086108 & -0.123971 & 5.651874 \\
\midrule
TOI-2076 \\
\midrule
b & -0.022501 & -0.022978 & 5569.717569 \\
\midrule
TRAPPIST-1 \\
\midrule
d & -0.088870 & -0.172715 & 7.576412 \\
e & -0.049611 & -0.050494 & 187.930194 \\
f & -0.011197 & -0.014690 & 42.653200 \\
\bottomrule
\end{tabular}
\caption{Identified potential secular resonances, respective values of the nodal and spin precession frequencies, and subsequent inclination-damping timescales.}
\label{tab:resonance_table}
\end{table*}

To determine whether resonance-driven obliquity is possible for a planet, we first must identify any secular spin-orbit resonances. These secular resonances are matches between a nodal precession rate in the planetary system and the spin axis precession of one of the system's planets, which causes a greater obliquity in the planet. We take a sample of ten multi-planet systems with four or more planets. Each of the planets in these systems have $R_{Earth} < R_p < 5.5R_{Earth}$ (with the exception of three planets in the TRAPPIST-1 system which are smaller than Earth). We slightly loosened out $5R_{Earth}$ upper bound for planets considered Neptune-like, and on the lower end we also include the few known Earth-sized planets. Given the assumption that the planets in these systems are despun, Super-Earths, which would otherwise have spin periods too rapid to maintain ring-moon cycles, may now fall within the "Boomerang" or "Torque-Dependent" regimes. These planets are thus included in the selection.

To confirm the planets in each system have been despun, we repeat the previous calculation of despinning timescales using Equation~\ref{eq:1}, but now using $Q = 100$ and $k_2 = 0.3$ for Super-Earths and smaller planets. This confirms that all planets in all systems have a despinning timescale of less than 1 Gyr, and thus we can assume synchronous rotations and calculate precessions accordingly.

The nodal precession rates of a system of $N$ planets can be calculated by determining the eigenvalues of an $ j \times k$ matrix $B$ with components \citep{murray1999}
\begin{gather}
    B_{jj} = -n_j \frac{1}{4} \frac{M_k}{M_{star} + M_j} \alpha_{jk} \bar \alpha_{jk} b^{(1)}_{3/2}(\alpha_{jk}) \\ 
    B_{jk} = +n_j \frac{1}{4} \frac{M_k}{M_{star} + M_j} \alpha_{jk} \bar \alpha_{jk} b^{(1)}_{3/2}(\alpha_{jk})
\end{gather}
where $j$ denotes the perturbing planet and $k$ denotes the planet being perturbed. Here $\alpha_{jk}$, $\bar \alpha_{jk}$, and $b^{(1)}_{3/2}(\alpha)$ are given by
\begin{gather}
\alpha_{jk} = 
\begin{cases}
    a_k/a_j & a_j > a_k \\
    a_j/a_k & a_j < a_k
\end{cases} \\
\bar \alpha_{jk} = 
\begin{cases}
    1 & a_j > a_k \\
    a_j/a_k & a_j < a_k
\end{cases} \\
b^{(1)}_{3/2}(\alpha) = \frac{1}{\pi} \int^{2\pi}_0 \frac{\cos{\psi}d\psi}{(1 - 2\alpha\cos{\psi} + \alpha^2)^{3/2}}
\end{gather}

Upon calculating the nodal precession frequencies, we also acquire a vector of $N$ values, each usually dominated by one of the system's planets. The largest of these values marks the planet contributing most to that particular precession.

The spin precession constant of a planet is given by \citep{winn2005}
\begin{equation}
    \dot \psi =  \frac{3}{2} \left( \frac{C - A}{C} \right) \left( \frac{n^2}{\omega} \right)
\label{prec_const}    
\end{equation}
where $C$ and $A$ are the principal moments of inertia and $\omega$ is the spin frequency. For despun planets, $n = \omega$. 

In general, for a non-rigid body that responds to tidal and centrifugal forces, the moments of inertia would depend both on the tidal field of the star and the spin of the planet. While zero-obliquity synchronous planets on circular orbits do not experience any variable forces, planets with obliquity do feel time-variable tides from the parent star. As the contributions of the rotation and stellar tides are of the same order \citep{murray1999}, here we will use only the rotational deformation to estimate the oblateness and precession rate of the planet. 

In general, the oblateness of a deformable rotating planet is proportional to the square of its spin rate, assuming its density and internal structure are constant. While the oblateness moment $J_2$ is a function of a planet's normalized moment of inertia $C/MR^2$, it does not depend on the tidal Love number, despite some confusion in the literature (see Appendix \ref{app}). Here we simply calculate the spin precession constant for each exoplanet by comparing it to either Neptune or Earth such that
\begin{equation}
    \dot \psi_{p} = 
    \begin{cases}
        \frac{3}{2} \left( \frac{C_N - A_N}{C_N} \right) \left( \frac{n_{p}}{\omega_{N}} \right)^2 n_{p} & R_{p} > 2R_{E} \\
        \frac{3}{2} \left( \frac{C_E - A_E}{C_E} \right) \left( \frac{n_{p}}{\omega_{E}} \right)^2 n_{p} & R_{p} < 2R_{E}
    \end{cases}
\end{equation}
where N and E denote Neptune and Earth, respectively.

To identify secular resonances in these systems, we need two things: firstly, $ g_i \leq \dot \psi_p \leq 2g_i$, where $g_i$ is one of the system's nodal precessions. This allows for secular spin-orbit resonances for planetary obliquities $\leq 60^{\circ}$. We limit ourselves to spin precession to orbit precession frequency ratio $<2$ and the corresponding resonant planetary obliquity below $60^{\circ}$, as the spin precession rate of a planet with obliquity $\theta$ is proportional to the precession constant (Eq \ref{prec_const}) times $\cos(\theta)$.

Our choice to limit ourselves to planets with resonant obliquities below 60$\circ$ (and therefore to spin-to-orbit precession period ratios between 0.5 and 1) is somewhat arbitrary, but does have a physical motivation. Solid triaxial satellites cannot maintain Cassini states with obliquities above 58$\circ$ \citep{bel72, gladman96, cuk16} and while this instability is not applicable to gas giants, it may have some applicability to Earth like planets (and by extension, possibly super-Earths), which do not reshape due to outside forces on timescales of a day (see Appendix \ref{app}). Prior simulations of stable high-obliquity spin-orbit resonances, such as \citet{Millholland2024}, typically integrated the planet’s precession but not the full rotation, potentially missing some of the instabilities. Note that we expect our planets to have largely damped their librations around the resonant obliquity, avoiding large scale periodic excursions to very high obliquities typical of undamped systems \citep{lu22}. 

Secondly, in the normal mode eigenvector associated with frequency $g_i$ (which is just a unit vector, as we do not know the real values of inclination), the value corresponding to the perturbed planet must be greater than 0.01. This is a relative measure of the influence of the normal mode on the perturbed planet, and the actual dynamical strength of the resonance also depends on the inclination associated with the normal mode. Inclinations to the invariable plane are not directly measurable (unless precession is detected), but they must be low ($<1^{\circ}$) in multi-transit systems. Such an inclination would result in a corresponding eigenvector component of less than 0.01$^\circ$. In particular, Kepler systems are likely to have inclinations on the order of 0.1$^\circ$, which would yield values well below 0.01$^\circ$. Therefore, we expect that any contributions of the eigenmode to the perturbed planet's inclination below this 0.01 threshold to be $< 0.01^{\circ}$ in absolute terms and therefore relatively minor.

Table~\ref{tab:systems_table} provides the list of example planetary systems, their planets, and basic planetary parameters. For each of these systems, we calculate the nodal precession frequencies and the spin precession frequencies, identifying any potential secular resonances. The total list of identified resonances is given in Table \ref{tab:resonance_table}.

\subsection{Equivalent Ring Thickness}

We next repeat the calculations for ring-moon lifetimes and the subsequent equivalent thickness of a 1-Gyr lifetime ring as done in Section~\ref{sec:ring_life}. The right plot of Figure~\ref{fig:ring_depths} shows the distribution of ring equivalent thickness for both Super-Earths and Neptune-like planets. Notably, fewer planets meet the ring thickness threshold in this case, with Super-Earths having especially low equivalent ring thickness. However, it is worth noting that regardless, many planets (including those in potential resonances) still meet or exceed the threshold. Note that, due to smaller tidal Qs assumed for Earth-type planets, satellite tidal lifetimes are orders of magnitude shorter, and rings are present much of the time.

\subsection{Inclination Damping due to Tides}

Obliquity tides typically act to damp the obliquity of synchronous rotators (such as moons), sometimes on short timescales \citep{Chen14}. However, when the obliquity of the orbiting body is forced by a spin-orbit interactions, then the obliquity tides act to decrease the inclination that is forcing this obliquity \citep{chyba1989, Fabrycky07, Chen16}. Planets experiencing significant damping of inclination due to obliquity tides will eventually lose their tilt, causing the rings to appear edge-on. On the other hand, however, slow rates of change in inclination lead to long system-flattening timescales, and thus it is more likely for rings to appear at an angle. With the potential secular resonances identified, we can analyze the change in inclination for each planet using \citep[modified from][]{chyba1989}

\begin{equation}
    \frac{di}{dt} = -\frac{3}{2}\left( \frac{\sin^2 \theta}{\tan i \left( \frac{M_{perturb}}{M_p} \right) \sqrt{\frac{a_{perturb}}{a_p}}} \right) \left( \frac{k_2}{Q_p} \right) \left( \frac{M_s}{M_p} \right) \left( \frac{R_p}{a_p} \right)^5 n
    \label{eq:17}
\end{equation}

where we assume $i = 0.1$ and $\theta$ is the inverse cosine of the ratio between the nodal and spin precessions. $M_{perturb}$ denotes the mass of the ``perturbing planet'', which is identified as the planet having the largest term in the inclination eigenvector corresponding to the secular frequency of the spin-orbit resonance. The idea here is that we are approximating the total angular momentum of this secular eigenmode by that of the planet most closely associated with the eigenmode. $M_p$ and $m_s$ denote the masses of the planet being perturbed (to which the spin precession in the potential resonance belongs) and the host star mass, respectively. $k_2/Q_p$ is assumed to be
\begin{equation}
\frac{k_p}{Q_p} = 
\begin{cases}
    \frac{0.1}{10000} & r_p > 2r_{Earth} \\
    \frac{0.3}{100} & r_p \leq 2 r_{Earth}
\end{cases} \\
\end{equation}

based upon the respective values for Earth and Neptune. All planets with potential resonances in the sample have an inclination-damping timescale greater than 5 Gyr, and thus it can be assumed that, if these systems were caught in spin-orbit resonances, the inclinations would not have been tidally damped. Note that an exception to this is Kepler-11, which has a known age of 8.5 Gyr \citep{lissauer2013}. However, the inclination-damping timescales found for planets in potential resonances in this system are still greater than the system's age, and thus the assumption still holds. Given this, any rings around planets in these potential spin-orbit resonances would not be edge-on, and therefore could potentially be observed during transits.

\section{Geometry of Transiting Exorings}

Transit photometry in particular has shown promise for detecting exoplanetary rings, as ring systems would both increase the transit depth and scatter light from the planet's host star \citep{barnes2004}. Increasing the transit depth can result in what appears as an abnormally large radius and, thus, a low density. Such an inflation of radius due to transiting ring systems around a planet has been discussed as a potential source of "super puffs," or planets with $\rho_p < 0.3$ g/cm$^3$, such as HIP-41378 f \citep{lu2024}. While the confusion between super-puffs and ringed planets is possible, advanced modeling of the transits can distinguish between these two cases. In this section we discuss the observability of rings around exoplanets in both singular and multi-planet systems.

\subsection{Calculating Ring Cross-Section For Non-Despun Planets}

A ring system viewed edge-on will not contribute significantly to the lightcurve of a transiting planet, and therefore in order for rings to be observable, the planet they orbit must have a significant obliquity such that the rings sit at an angle relative to the observer. For non-despun Neptune-like planets we assume that their spin poles are randomly distributed on the celestial sphere, based on moderate obliquity of Neptune and very large obliquity of Uranus. Our coordinate system uses the transiting planet's orbit as the reference plane, with the direction to Earth being the zero position in longitude. Thus, we select a random uniform distribution of longitude angles, $\gamma$, choosing from ranges of $0 \leq \gamma \leq 2\pi$. Our selection of latitude angles varies between non-despun planets and despun planets in potential spin-orbit resonances. For non-despun planets, we take a uniform distribution of obliquities $\sin{\theta}$ between -1 and 1, then take the latitude angle as $90 - \theta$. For despun planets in potential spin-orbit resonances, we utilize the obliquity  $\theta$ as used in Equation~\ref{eq:17} instead of a uniform distribution. Using one hundred sets of these two angles (either one hundred unique longitudes and latitudes for non-despun planets or one hundred longitudes paired with one latitude for the planets in potential spin-orbit resonances) to represent the spin axis, we then find the angle between that spin axis and the observation line from Earth, here defined to be the x-axis. 

We take the maximum area of the ring cross-section to be a circle with radius $R_{Ring} = a_{FRL}$. The surface area of the ring as seen from Earth, then, will be $\pi R_{Ring}^2 \cos{\alpha}$, where $\alpha$ is the angle between the spin axis and the observer. From this, we get the semimajor axis $a$ and semiminor axis $b$ of an ellipse, where the semimajor axis is always $a_{FRL}$. We are ignoring any gaps between the inner edge of the rings and the planet as they they are likely to have relatively small projected cross sections and would just add unnecessary complexity to the calculation.

\begin{figure}
\centering
\begin{tikzpicture}
    \filldraw[fill = black!20!white] (0,2) ellipse (3cm and 2cm);
    \filldraw[fill=white] (0,2) circle (1cm);
    \draw[thick, red, dashed] (0,2) -- (0,4) node[above]{\large b};
    \draw[thick, red, dashed] (0,2) -- (3,2) node[right]{\large a};
\end{tikzpicture}
\begin{tikzpicture}
    \usetikzlibrary {angles}
    \usetikzlibrary{quotes}
    
    \filldraw[fill = black!20!white] (2,2) ellipse (3cm and 0.8cm);
    \filldraw[fill = white] (2,2) circle (2cm);
    \draw (2,2) ellipse (3cm and 0.8cm);
    \draw (2,2) -- (0.16,2.8);
    \draw (2,2) -- (0.16, 1.2);
    \draw (2,2) -- (3.84, 2.8);
    \draw (2,2) -- (3.84, 1.2);
    \draw[thick, red, dashed] (2,2) -- (2,2.8) node[above]{\large b};
    \draw[thick, red, dashed] (2,2) -- (5,2) node[right]{\large a};
    \draw[thick, dashed](0.16, 2.8) -- (3.84, 2.8);
    \draw[thick, dashed] (3.84, 1.2) -- (0.16, 1.2);
    \path (current bounding box.north) ++ (0,0.5cm);
    \draw (2, 2.2) coordinate (A) -- (2,2) coordinate (B)
         -- (3.84,2.8) coordinate (C)
           pic ["$\sigma$", draw, thick,angle radius=0.75cm] {angle = C--B--A};
    \draw (0.16, 1.2) coordinate (A) -- (2,2) coordinate (B)
         -- (0.16, 2.8) coordinate (C)
           pic ["$\phi$", draw, thick,angle radius=0.75cm] {angle = C--B--A};
\end{tikzpicture}
\caption{Diagram of the two types of ring-planet cross-sections. The upper image shows the case where $b > R_p$, while the lower shows $b < R_p$. In the second case, the overlap between the ring and planet is taken as two straight lines across the planet area.}
\label{fig:diagram}
\end{figure}
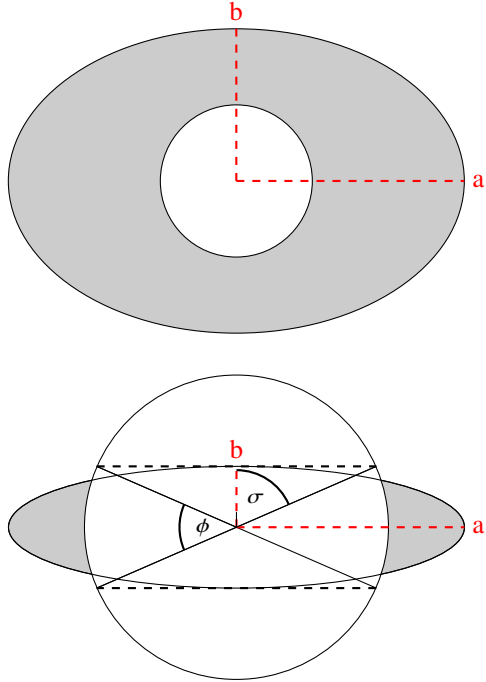

The cross sections can be addressed in two ways, as shown in Figure \ref{fig:diagram}. The first case, where $b > R_p$, yields a cross-section where the ring entirely encompasses the planet. Here, the calculation of the portion of the cross-section attributed to the ring is
\begin{equation}
    A_{Ring} = \pi(a_{FRL}b - R_p^2)
\end{equation}
where $R_p$ is the radius of the planet.
In the second case, $b < R_P$. Here, we assume the intersections between the ring and the planet to be two straight lines at distances $b$ and $-b$ from the center of the circle such that the overlapping space can be approximately split into two isosceles triangles and two circular sectors. The triangles have leg lengths of $R_p$ and heights of $b$. The circular sectors have radii of $R_p$. As such, the area of overlap between the planet's cross-section and the ring's can be calculated as
\begin{equation}
    A_{overlap} = 2bR_p\sin{\sigma} + R_p^2\phi
\end{equation}

where $\sigma$ is the angle between the semiminor axis $b$ and the line between the center of the planet's cross-section and the upper intersection point between the planet and ring cross sections. Likewise, $\phi$ represents the angle between the upper and lower intersection points of the cross sections. Here, then, the total ring-contributed cross-section is given by
\begin{equation}
    A_{Ring} = \pi a_{FRL}b - A_{overlap}
\end{equation}
and, in both cases, the total cross section of both ring and planet is given by
\begin{equation}
    A_{Total} = \pi R_p^2 + A_{Ring}
\end{equation}
Figure~\ref{fig:diagram} shows the intersection of the cross sections in both cases.

\begin{figure*}
    \centering
    \includegraphics[width=\textwidth]{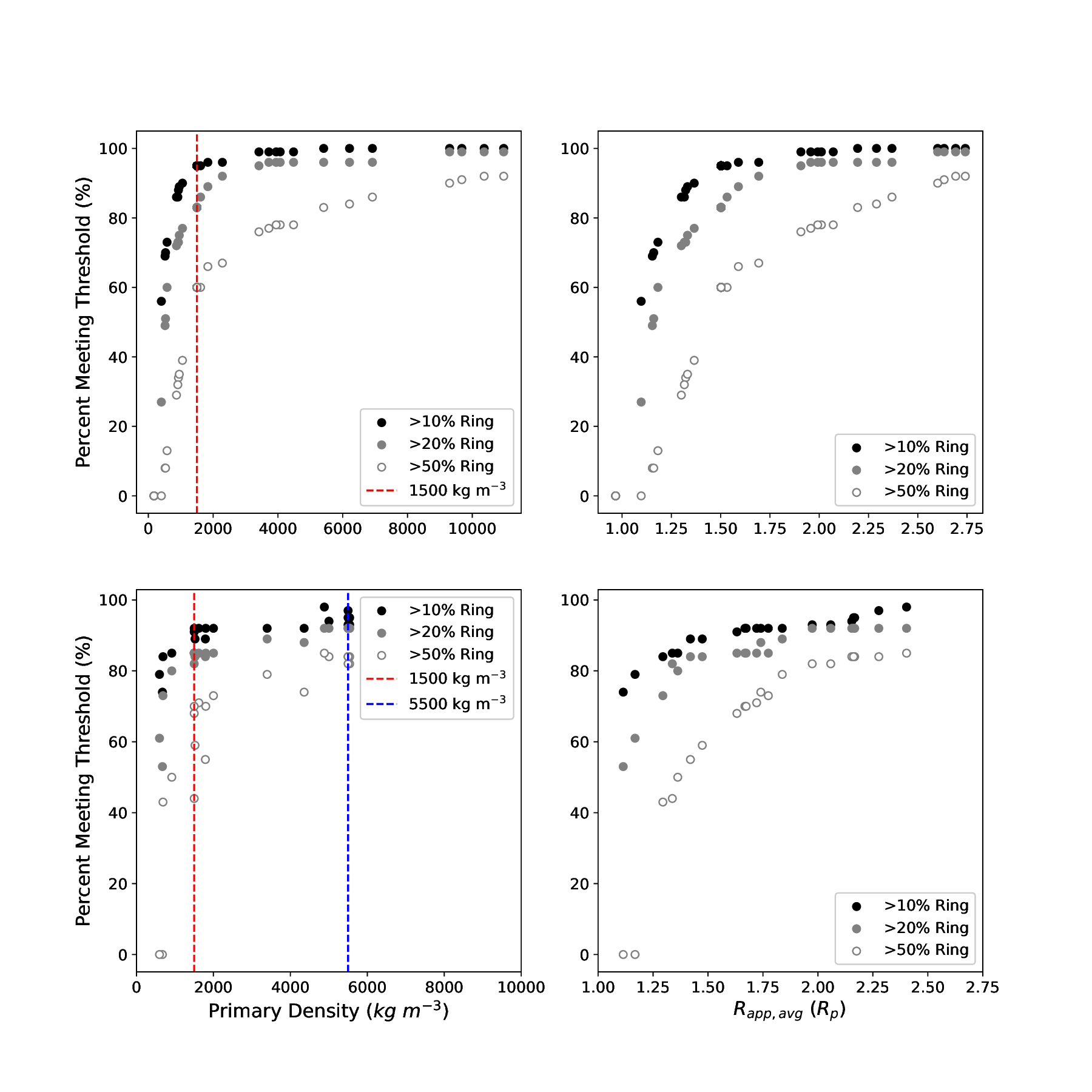}
    \caption{The percentage of planets meeting various cross-section ring contribution thresholds (left) and, subsequently, the apparent radii for planets with those cross-sections (right). The top plots represent the finalized sample of non-despun planets and the bottom plots represent despun planets, specifically those in potential spin-orbit resonances. Note that the plot of threshold-meeting cases for non-despun planets also marks the line $\rho = 1500$ kg m$^{-3}$, and for planets in potential resonances marks both that line and $\rho = 5500$ kg m$^{-3}$; these are the assumed density for planets without defined masses.}
    \label{fig:thresholds_and_appradii}
\end{figure*}

By finding the ratio $A_{Ring}/A_{Total}$, we can identify the percentage of the cross-section contributed by the ring in each case and, thus, the percentage of cases for each planet that yield a ring contribution above a certain threshold. For our purposes, we establish three such thresholds: 10$\%$, 20$\%$, and 50$\%$. Cases which meet or exceed these thresholds have significant contributions by the rings to their cross-sections, and thus it could be plausible to see rings in the transits of these planets. Figure ~\ref{fig:thresholds_and_appradii} shows the percentages of cross-sections beyond each of these thresholds for both non-despun planets and despun planets in potential spin-orbit resonances.

\subsection{Super-Puff Exoplanets and Exorings}

A notable question that has emerged in recent years is the origin of "super-puff" planets, where $\rho < 300$~kg~m$^{-3}$. For some of these, their extremely low densities have been theorized to be due to opaque ring systems. A ring present in a transit lightcurve can increase the depth of the transit, subsequently leading to a larger calculated radius and a lower calculated density \citep{zuluaga2015, piro20, Saillenfest23, lu2024} Such a ring system has been thus proposed for very low density ``super-puff'' planets \citep{Masuda14, Lopez14, Lee16} like HIP-41378 f \citep{akinsanmi2020, lu2024}. To explore this, we calculate the ``apparent'' radius of the planet by adopting the (incorrect) assumption that the entire cross-section is due to the planet and re-calculating the planet's implied radius accordingly. The average apparent radius, set as the mean apparent radius among all cross-sections of a given planet, reaches up to approximately 2.75 $R_{p}$. The overall range of differences between apparent and true radii is shown in Figure~\ref{fig:thresholds_and_appradii}. Generally, the apparent radius increases as the number of spin axis orientations meeting the ring cross-section contribution thresholds increases.

\section{Conclusions}

In this paper, we have explored the plausibility of ring-moon cycles, originally proposed for planets in our Solar System \citep{HM17}, to be sources of late forming and/or long-lived rings that may be detectable around transiting exoplanets. We have identified two distinct dynamical regimes in which ring-moon cycles may operate and be detectable. One involves non-despun Neptune-like planets, and the other members of multi-planet systems that are caught in spin-orbit resonances. 

Both non-despun and despun pathways show promise that ring-moon cycles can plausibly be taking place in a number of exoplanetary systems. We find that ring-moon cycles around non-despun Neptune-like planets tend to form rings with greater equivalent thickness that can contribute more to transiting lightcurves, but are often present only about 20\% of the time. Tidally despun planets in spin-orbit resonances are expected to have lower surface mass densities for rings formed from-ring moon cycles. However, as many of them are high-density super-Earths with large Roche limits, their rings are expected to be larger relative to the planet, and rings may be present more than half of the time. Apart from changing the profile of the transit, the presence of rings can given an illusion of a larger planetary radius, resulting in an anomalously low measured planetary density \citep[cf. ][]{lu2024}. 

While the transit profile of the rings can be confounded with certain properties of the star such as limb darkening, multi-planet transiting systems may break this degeneracy. Even if multiple planets have rings it is unlikely that all of them would be in spin-orbit resonances that can maintain non-zero obliquity, so it should be possible to compare transits of ringless and ringed planets (if any) in the same system, making it easier to detect the rings. Alternatively, advanced modeling of transits \citep{cassese24, dholakia25, Price25} can make it possible to distinguish between rings and extended atmosphere even without comparison to other transiting planets in the system.

\section*{Acknowledgements}

We wish to thank the NSF Research Experience for Undergraduates (REU) program at the SETI Institute and private donors to the program for making this research possible, and Matthew Tiscareno and Nisha McFarland for all their help during this program. We thank Matt Tiscareno, Doug Caldwell, and Darin Ragozzine for very helpful comments on the manuscript. 

%%%%%%%%%%%%%%%%%%%%%%%%%%%%%%%%%%%%%%%%%%%%%%%%%%
\section*{Data Availability}

Analysis scripts and other supporting data are available from the corresponding author upon
reasonable request.



\bibliographystyle{mnras}
\bibliography{example} % if your bibtex file is called example.bib

%%%%%%%%%%%%%%%%% APPENDICES %%%%%%%%%%%%%%%%%%%%%
\appendix

\section{Oblateness Moment $J_2$ of a Rotating Planet}\label{app}

While in this paper we have simply scaled the oblateness of the exoplanet to that of Earth or Neptune using the planet's spin period, we have found a large variation in how the oblateness moment $J_2$ of a rotating planet is calculated in the literature. In particular, such calculations often include the tidal Love number of the planet, which is relevant for a planet's response to time-dependent perturbations, while the rotational distortion is time-independent. For this reason we include an explicit version of the relationship between $J_2$ and the rotational parameter $q$, derived from \citet{murray1999}. This relation is not new or original but may be useful to the community. 

\citet{murray1999} in their Section 4.5 present the relationship between flattening $f$, oblateness moment $J_2$ and rotational parameter $q$ as:

\begin{equation}
f=\frac{3}{2}J_2+\frac{1}{2}q
\label{app:f}    
\end{equation}
\noindent where 
\begin{equation}
J_2=\frac{C-A}{M R^2}
\label{app:j2}
\end{equation}
with $C$ and $A$ being polar and equatorial moments of inertia, and 
\begin{equation}
q=\frac{\omega^2 R^3}{GM}
\label{app:q}
\end{equation}
\noindent where $\omega$ is the spin period of the planet. 
If we assume that the planet is in hydrostatic equilibrium, the Darwin-Radau relation may be used \citep{Cook1980}:
\begin{equation}
\frac{J_2}{f} = - \frac{3}{10} + \frac{5}{2}\bar{C} - \frac{15}{8} \bar{C}^2 = \frac{1}{\Phi(\bar{C})}
\label{app:dr}
\end{equation}
where we newly define $\Phi$ as a function of $\bar{C}$, the normalized moment of inertia (Section 2.1). Combining Equations \ref{app:f} and \ref{app:dr}, we get:
\begin{equation}
J_2 = \frac{q}{2 \Phi(\bar{C)}-3}
\label{derived}
\end{equation}
For a uniform density planet, $J_2=q/2$. For Earth $\bar{C}=0.3307$ \citep{williams94}, so $J_2= q/3.217=1.077 \times 10^{-3}$, as opposed to measured $J_2=1.083 \times 10^{-3}$ \citep{murray1999}. For Saturn $\bar{C}=0.219$ \citep{Mankovich23}, so $J_2=q/9.69=0.01637$ as opposed to measured $J_2=0.01630$ \citep{murray1999}. We give the above numbers to illustrate that Eq. \ref{app:j2} can be used to estimate $J_2$ to a precision that is fully adequate for purposes of exoplanet science, without involving the $k_2$ Love number. 

A different purported relationship 
\begin{equation}
J_2=\frac{k_2 q}{3}
\label{wrong}
\end{equation}
and has been used somewhat incorrectly by otherwise comprehensive and insightful exoplanet spin-orbit resonance papers \citep[e.g.][]{Millholland19, Millholland2024, lu2024}. This relation is exactly true for a uniform-density fluid planet, and the inverse relation has been used to estimate the Love number of gas giants from $J_2$, as both $J_2$ and $k_2$ measure central concentration of mass \citep{yoder95}. The principal modern source of this relation is  \citet{Ragozzine09}, who independently derived tidal and rotational potentials, but appear not to have fully differentiated between time-variable and permanent distortion. As Earth is not fluid on tidal timescales, Eq. \ref{wrong} is off by a factor of a few. Saturn's $k_2$ is augmented by its fast rotation, and $J_2$ from Eq. \ref{wrong} is overestimated; in general the relationship between $k_2$ and $J_2$ is complex even for fluid planets \citep{consorzi23}. 

\citet{Neron97} has also been cited as the source of Eq. \ref{wrong}, but this appears to be erroneous. Additionally, attribution to \citet{Sterne39} has appeared in the literature, but this appears to be an unfortunate confusion, as his $k_2$ is stated to be the same as $k$ in \citet{Russell28}, which is simply $(3/2) J_2$ in modern notation. This is clear from comparison of the first equation in \citet{Russell28} to Eqs. 4.110 and 6.249 in \citet{murray1999}. We hope that the community adopts a more consistent approach to estimating exoplanet oblateness in the future.

%%%%%%%%%%%%%%%%%%%%%%%%%%%%%%%%%%%%%%%%%%%%%%%%%%

% Don't change these lines
\bsp	% typesetting comment
\label{lastpage}
\end{document}